\documentclass[epj,aps,pra,twocolumn,showpacs,preprintnumbers,amsmath,amssymb,footinbib]{svjour}

\usepackage{float}
\usepackage{epsfig,amsmath} 
\usepackage{hepnames,hepunits}
\usepackage{xcolor}
\def\cascade{{\sc Cascade}}

\def\lsim{\mathrel{\rlap{\lower4pt\hbox{\hskip1pt$\sim$}}
    \raise1pt\hbox{$<$}}}                
\def\gsim{\mathrel{\rlap{\lower4pt\hbox{\hskip1pt$\sim$}}
    \raise1pt\hbox{$>$}}}                

\newcommand{\alphas}{\ensuremath{\alpha_\mathrm{s}}}

\newcommand{\PBM}{PB}

\newenvironment{tolerant}[1]{\par\tolerance=#1\relax}{ \par }

\begin{document}

\headnote{
\begin{flushright}
\fontfamily{pcr}\fontseries{m}\fontsize{12}{16}\selectfont
DESY 21-014\\
2 Feb 2021
\end{flushright}
 }
\title{\boldmath  Determination of collinear and TMD photon densities using the Parton Branching method}

\author{
H.~Jung \inst{1} \and 
S.~Taheri~Monfared \inst{1} \and  
T.~Wening \inst{2} 
}

\institute{Deutsches Elektronen-Synchrotron, D-22607 Hamburg,
\and
II. Institut f\"ur Theoretische Physik, Universit\"at Hamburg}

\abstract{We present the first determination of transverse momentum dependent (TMD) photon densities with the Parton Branching method. The photon distribution is generated perturbatively without intrinsic photon component.
The input parameters for quarks and gluons are determined from fits to precision measurements of deep inelastic scattering cross sections at HERA.
The TMD densities are used to predict the mass and transverse momentum spectra of very high mass lepton pairs from both Drell-Yan production and Photon-Initiated lepton processes at the LHC.}

\maketitle
\flushbottom
\section{Introduction}
The  Parton Branching (PB) evolution method \cite{Hautmann:2017xtx,Hautmann:2017fcj} has been applied to evolve both collinear and transverse momentum dependent parton distributions (TMDs) from a small to large scale using the DGLAP evolution equation. An important feature of this method is that it gives a solution which is fully exclusive, and therefore allowing for a determination of the TMD parton density. The PB method applies the unitarity formulation of the QCD evolution equation and has shown to be valid for leading-order (LO), next-to-LO (NLO) and next-to-NLO (NNLO). 

 Given that $\alpha_s^2\sim\alpha$ over a wide range of scales, it becomes necessary to include also the corresponding electroweak (EW) corrections in the evolution, which can exceed the few percent level and become quantitatively very important for an accurate prediction \cite{Bauer:2017bnh,Bauer:2017isx}. So far, QED corrections have been taken into account for observables involving collinear parton distribution functions (PDFs) \cite{Roth:2004ti,Martin:2004dh,Ball:2013hta,deFlorian:2015ujt,deFlorian:2016gvk,deFlorian:2018wcj,Giuli:2017oii,Harland-Lang:2019pla,Schmidt:2015zda,Sadykov:2014aua,Carrazza:2015dea,Harland-Lang:2016kog,Manohar:2016nzj,Manohar:2017eqh}. Such analyses performed up to NNLO in QCD and LO in QED show that the photon PDF contribution is not negligible and needs to be carefully studied for precise predictions at the LHC and even more for higher energies as the HE-LHC and FCC-hh, where a particularly important aim involves events with leptons in the final state.
  The first significant change in the evolution of parton distributions with QED corrections is the appearance of the photon density. In this context, it is necessary and timely to consider the QED contribution to the PB evolution and to extract the first photon TMD. Recent phenomenological studies of contributions from photon-initiated (PI) channels to lepton pair production based on the structure function calculation of the underlying process in proton-proton collisions are discussed in \cite{Harland-Lang:2020veo,Harland-Lang:2021zvr}.

In this report the determination of parton densities with QED corrections obtained using the PB method is presented together with applications of the obtained photon TMDs to high mass lepton pair production at LHC energies.

\section{Method description}
In the PB approach the complete evolution of the parton density including the full information of the kinematic of the evolution process is calculated. Soft gluon emission and transverse momentum recoils are expressed by 
introducing the soft gluon resolution scale $z_M$. The evolution without resolvable branching from $\mu_0^2$ to $\mu^2$ is treated via Sudakov form factors
\begin{equation}
\Delta_a(\mu^2,\mu_0^2)=\texttt{exp} \left (-\sum_b\int_{\mu_0^2}^{\mu^2}\frac{dq^{ 2}}{q^{ 2}} \int_0^{z_M}  dz~ z~ P_{ba} ^{(R)}(\alpha_s, z) \right ).
\end{equation}
Here, $z_M$ separates resolvable and non-resolvable branchings, $z$ is the longitudinal momentum fraction, $\alpha_s$ is the strong coupling and $P_{ba}^{(R)}$ represents the real emission splitting functions from flavour $a$ to $b$. In this approach the TMD evolution equations are written as 
\begin{eqnarray}
\label{evoleqforA1}
  && { {\cal A}}_a(x,k_t^2, \mu^2) 
 =  
 \Delta_a (  \mu^2  ) \ 
 { {\cal A}}_a(x,k_t^2,\mu^2_0) + \nonumber\\ 
 && \sum_b 
\int
{{d { q}^{\prime 2} } 
\over { { q}^{\prime 2} } }
 \ 
 {{d \phi}
 \over
 {2\pi}
 }
{
{\Delta_a (  \mu^2  )} 
 \over 
{\Delta_a (  { q}^{\prime 2}  
 ) }
}
\ \Theta(\mu^2-{ q}^{\prime 2}) \  
\Theta({ q}^{\prime 2} - \mu^2_0)\times
 \nonumber\\ 
&&  
\int_x^{z_M} {{dz}\over z} \;
P_{ab}^{(R)} (\alphas 
,z) 
\;{ {\cal A}}_b\left({x \over z}, k_t^{\prime 2} , 
{ q}^{\prime 2}\right)  
  \;\;  ,     
\end{eqnarray}
where $ { {\cal A}}_a(x,{ k_t^2}, \mu^2)$ is the  TMD distribution of flavour $a$, carrying the longitudinal momentum fraction $x$ of the hadron's momentum and transverse momentum ${k_t^2}$ at the evolution scale $\mu^2$. The transverse momentum is given by $k_t^{\prime}=|{\bf k}+(1-z) {\bf q}^\prime|$, where ${\bf q}^\prime$ is the rescaled transverse momentum vector of the emitted parton and $\phi$ is the azimuthal angel between ${\bf q}^{\prime}$ and ${\bf k}$. In the application of Eq. \ref{evoleqforA1} we consider the scale at which $\alpha_s$ is evaluated not necessarily equal to the evolution scale. We apply the choice enforced by an angular ordering of the emissions therefore ensuring quantum coherence of softly radiated partons. 
The first set of TMDs evolved with NLO QCD DGLAP splitting functions determined from a fit to HERA I+II precision measurements \cite{Abramowicz:2015mha} have been described in Ref. \cite{Martinez:2018jxt}. 

Here, we concentrate on QED corrections.
The LO-QED kernels are \cite{Roth:2004ti}
\begin{eqnarray}
P_{qq}&=&e_q^2~\frac{1+z^2}{[1-z]_+}+\frac{3}{2}~e_q^2~\delta(1-z)~,\nonumber\\
P_{q\gamma}&=&N~e_q^2~(z^2~+~(1-z)^2)~, \nonumber\\
P_{\gamma q}&=&e_q^2~\frac{1+(1-z)^2}{z}~,\nonumber\\
P_{\gamma\gamma}&=&-\frac{N}{3}~ \sum_q~e_q^2~\delta(1-z),
\end{eqnarray}
 where the sum $\sum_q$ only goes over all active flavours $N$ and we neglect leptonic contributions to $P_{\gamma\gamma}$. The resolvable kernel $P_{\gamma\gamma}^{(R)}$ vanishes, as $P_{\gamma\gamma}$ only contains a part proportional to the Dirac distribution.
Since the LO-QED splitting kernels depend on the electric charge of the quark, the evolution for up-type and down-type quarks is different. The momentum sum rule holds for the LO QED splitting kernels. A full standard model evolution equations, including a Sudakov form factor is discussed in  \cite{Bauer:2017bnh,Bauer:2017isx}.

The QED evolution is performed using the \PBM\ method, assuming the photon is generated dynamically only from photon radiation off the quarks (available in  the extended version of \verb+uPDFevolv+ \cite{updfvolv:2021}).

\section{Collinear photon density}
Several groups have determined collinear photon PDFs: The MRST \cite{Martin:2004dh} group used a parametrization for the photon PDF based on radiation off of “primordial” up and down quarks. In CT14  \cite{Schmidt:2015zda} a similar phenomenological model was adopted. The NNPDF group \cite{Ball:2013hta} and  xFitter Developers’ Team \cite{Giuli:2017oii} 
treated the photon PDF on the same footing as the quark and gluon PDFs. Within their approach, the photon PDF is parametrized at the starting scale. 

MMHT \cite{Harland-Lang:2019pla} provided the photon PDF separated into elastic (the photon component generated by coherent radiation from the proton as a whole) and inelastic contributions (the photon component generated from quarks), while the elastic component is less significant at higher $Q^2$ and negligible below $x \sim 0.2$. 

In the \PBM\ approach we generate the photon by perturbative radiation, without any intrinsic photon distribution.
 We constrain the QCD partons by a fit to HERA data in the ranges $3.5 < Q^2 < 50000$ GeV$^2$ and $4~ .~ 10^{-5} < x < 0.65$ at NLO in QCD, for $\alpha_s(M_Z)=0.118$, with LO QED evolution. We perform the evolution with Set~2 settings of Ref. \cite{Martinez:2018jxt}. The inclusion of a photon PDF has a negligible impact on other PDFs. The fits of both QCD set \cite{Martinez:2018jxt} and QCD+QED set with the same functional form for the gluon and quark distributions resulted in a similar $\chi^2/dof=1.21$. 
\begin{figure}[H]
\begin{center} 
\includegraphics[width=0.42\textwidth]{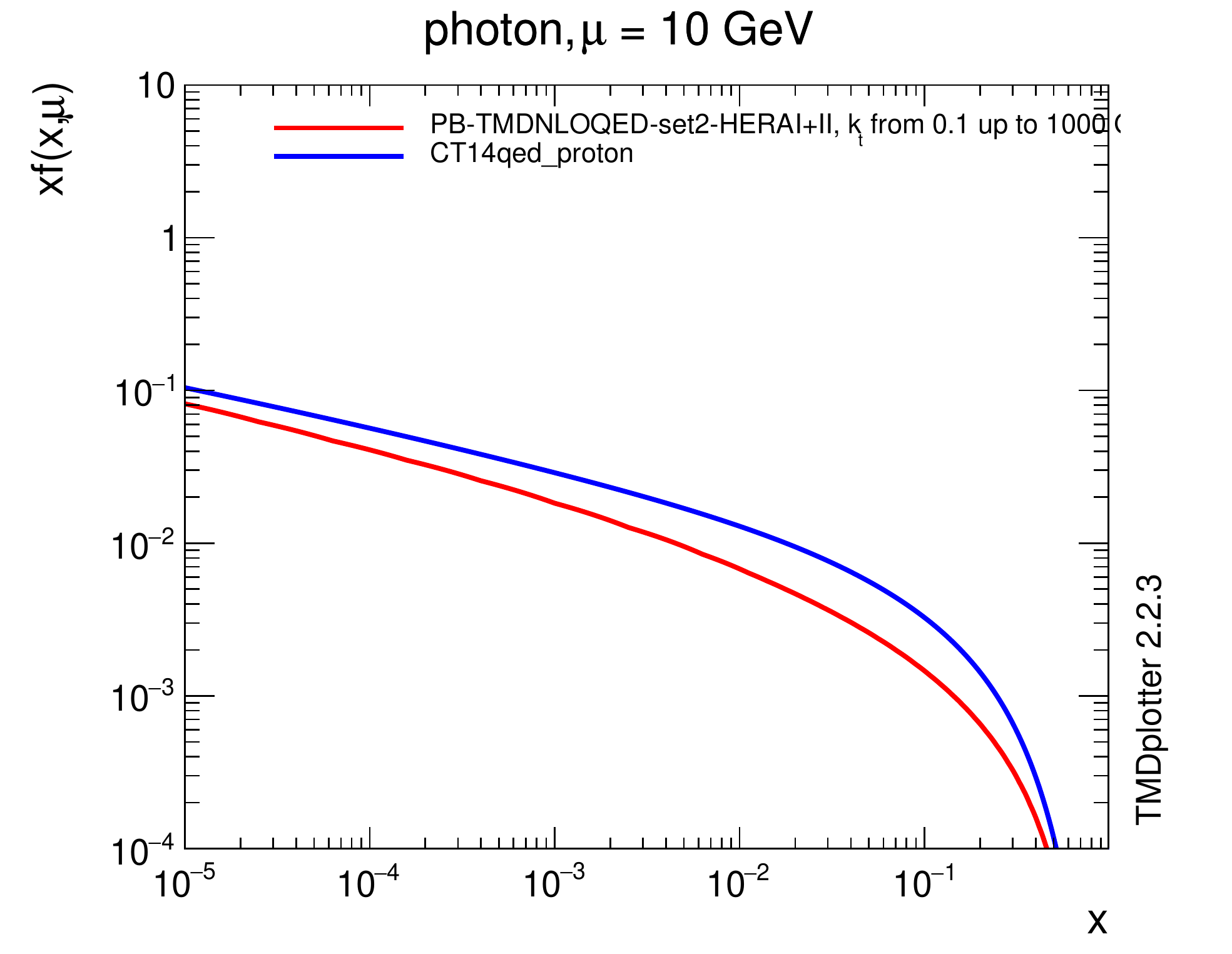}

\includegraphics[width=0.42\textwidth]{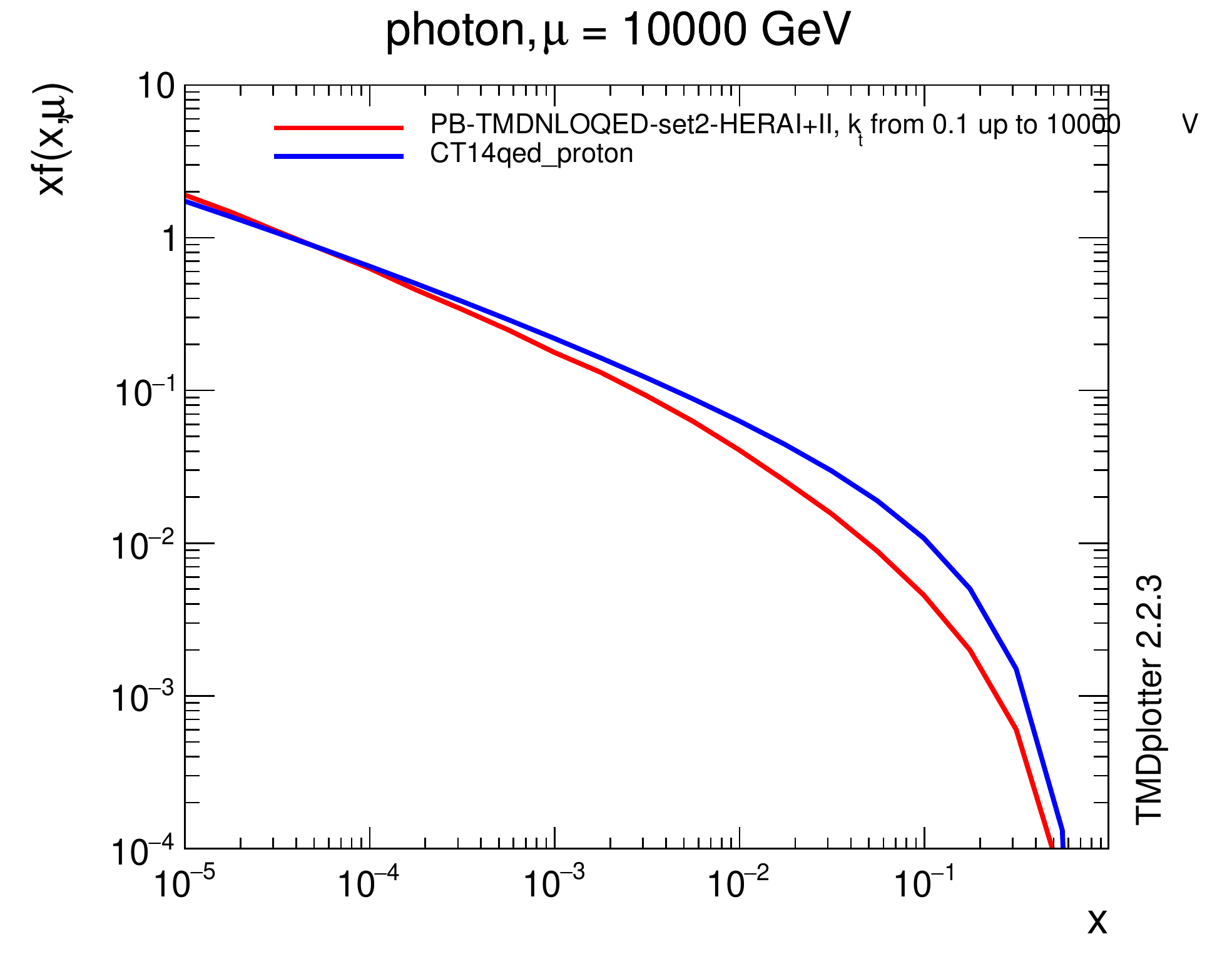}

 \caption{The photon PDF at $Q=10$ GeV and $Q=10^4$ GeV plotted versus $x$. CT14qed-proton is also shown for comparison. }\label{collinear-ph}
\end{center} 
\end{figure}
We have performed a benchmark test as in Ref. \cite{Carrazza:2015dea} by taking the same parametrization and initial scale in the FFN scheme with only four active quarks. With the same assumption of $\gamma(x,Q_0^2)=0$, an excellent   agreement is achieved for all flavors, photon and gluon  PDFs. 

In Fig. ~\ref{collinear-ph} we compare the collinear PB photon PDF with CT14qed at the scale of $Q=10$ GeV and $Q=10^4$ GeV. At large scale and small $x$ curves are very similar.

\section{TMD photon density}
TMD parton densities can be obtained within the \PBM\ method. The procedure for the determination of the TMD distributions of quarks and gluons is the same as described in Ref. \cite{Hautmann:2017xtx,Hautmann:2017fcj,Martinez:2018jxt}.
	
\begin{figure}[H]
\includegraphics[width=0.42\textwidth]{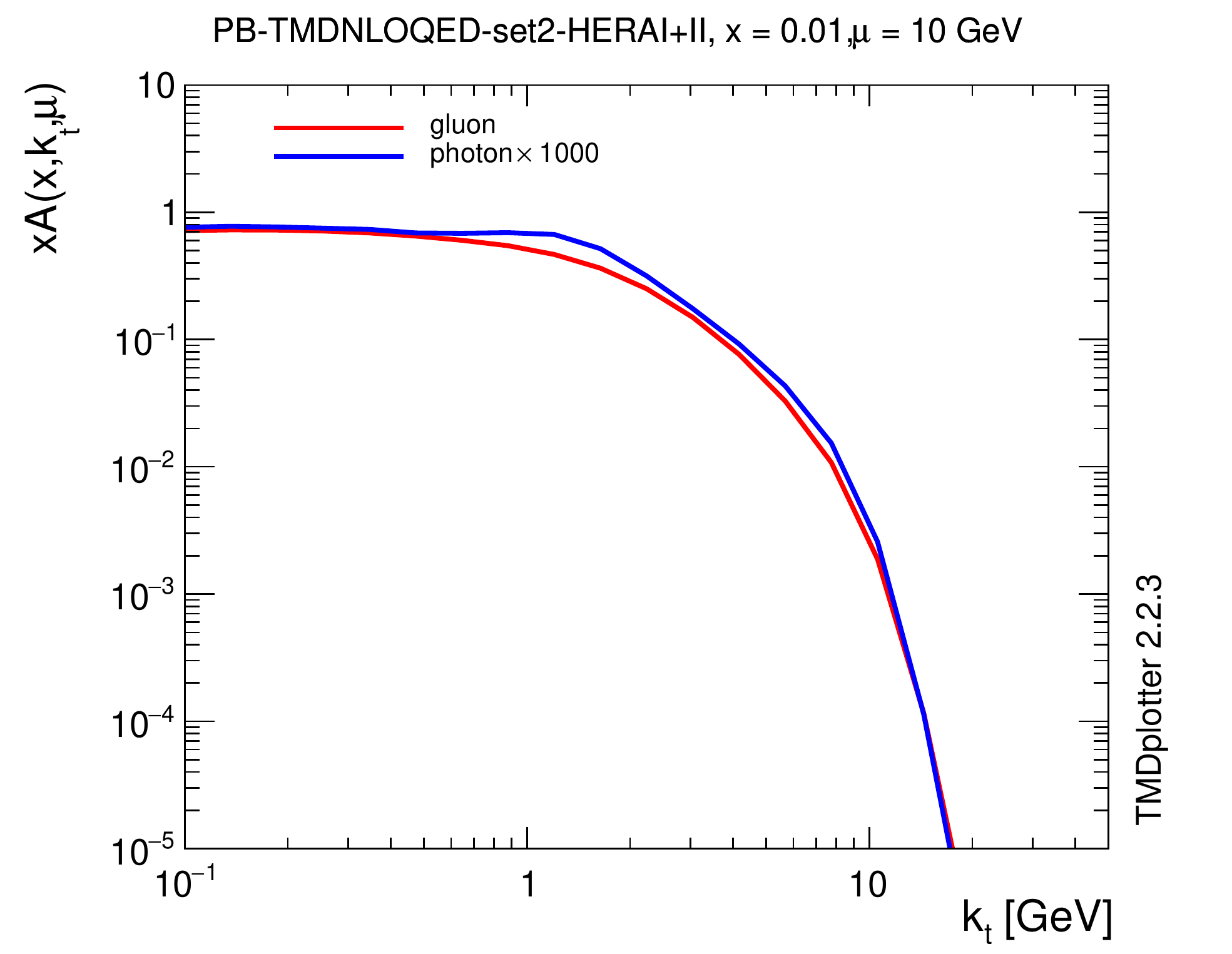}
\includegraphics[width=0.42\textwidth]{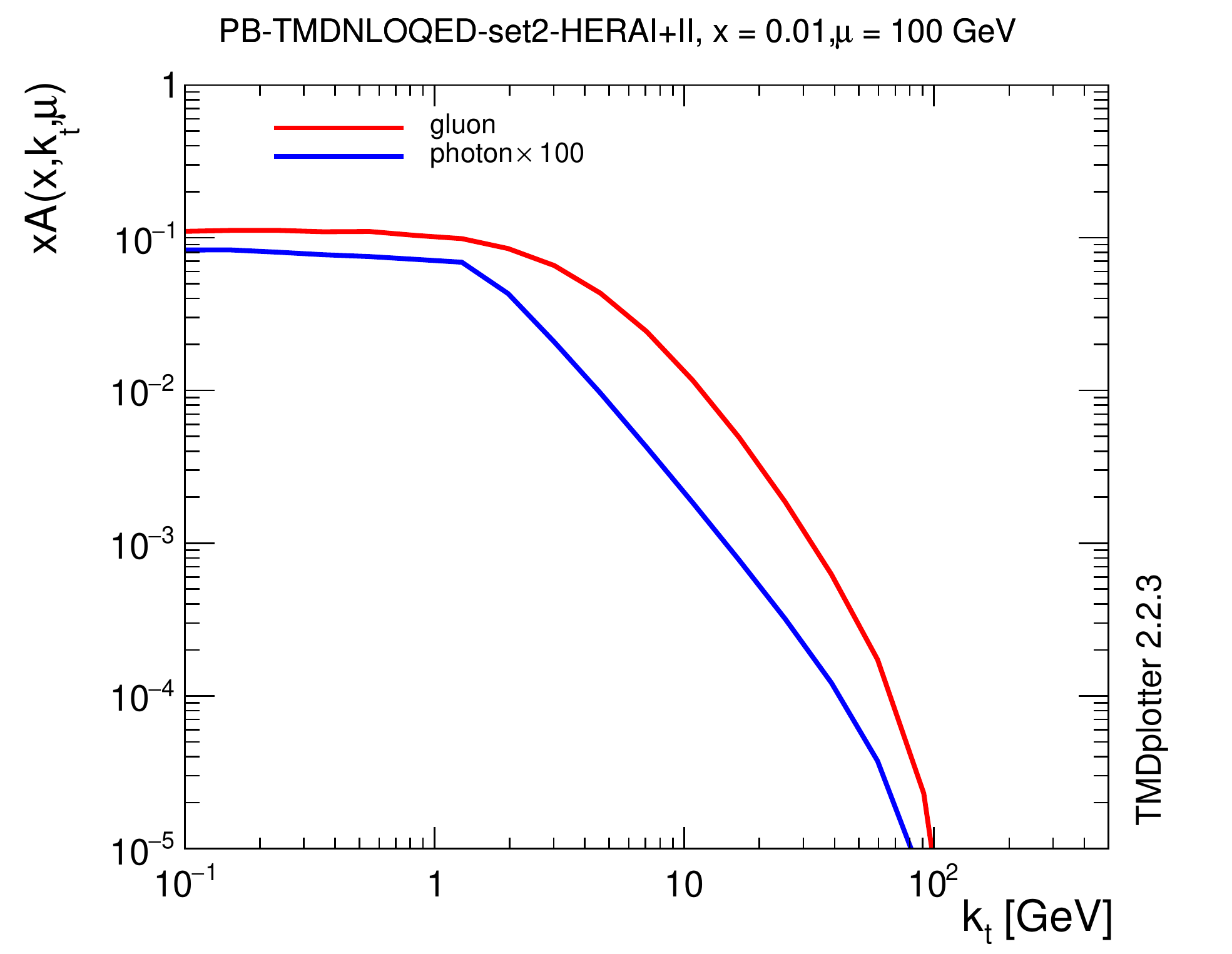}
\caption{Transverse Momentum Dependent photon and gluon densities at $x=0.01$ as a function of $k_t$ for different scales $\mu=10$ GeV and $\mu=100$ GeV.}\label{TMD-pho}
\end{figure}

Fig.~\ref{TMD-pho} shows the gluon and scaled photon TMD for  $\mu=10$~GeV and $\mu=100$~GeV. 
The shape of distribution at low $k_t$ ($k_t< 1$ GeV) is similar while at large $k_t$, especially at large scale ($\mu=100$ GeV), there is a difference coming from perturbative gluon-gluon splitting which has no correspondence in the photon case  (this effect appears in the second term in eq. (\ref{evoleqforA1})).

\section{Application to very high lepton pair production mass}
The CMS experiment \cite{Sirunyan:2018owv} has measured the production of pairs of muons over a wide range of  the dilepton invariant mass. Dilepton production in hadron-hadron collisions provides a unique tool for improving our understanding of hadronic structure and in particular for testing parton distributions. The contributions from PI lepton production ($\gamma \gamma \rightarrow l^{+}l^{-}$, with $l=e, \mu$) in hadron-hadron collisions are sizable at high invariant mass \cite{Harland-Lang:2016kog,Bourilkov:2016qum,Bourilkov:2016oet,Accomando:2016tah,Accomando:2016ehi}. 

\begin{figure}[H]
\begin{center} 
\includegraphics[width=0.5\textwidth]{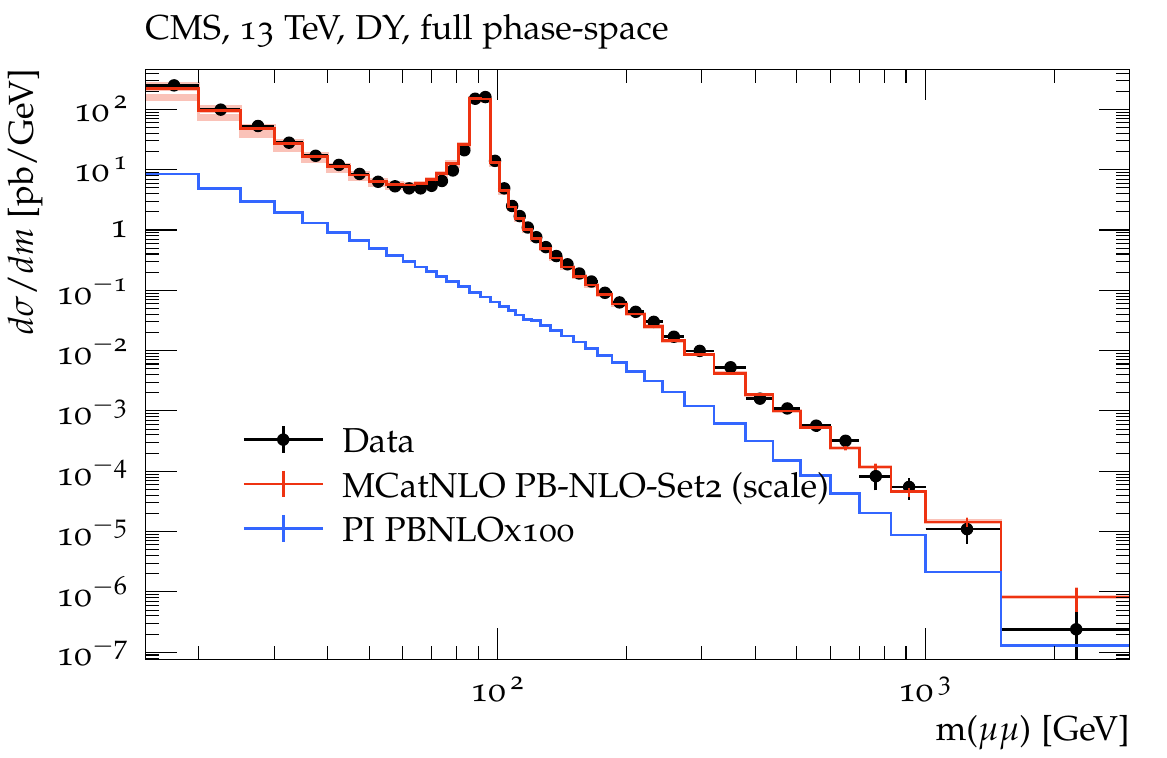}
  \caption{Dilepton high mass distribution production compared to predictions at QCD+QED using PB-TMDs in the full phase space. The scaled PI contribution is also shown.  }
\label{TMD_pdfs3}
\end{center}
\end{figure}

\begin{tolerant}{1500}
In Fig. \ref{TMD_pdfs3} we show the measured dilepton mass spectrum and compare it with the prediction of collinear NLO PB-QED (Set~2). The spectrum is rather well described with the NLO PB-QED prediction. 

{\sc MadGraph5\_aMC@NLO} \cite{Alwall:2014hca} is used to calculate the dilepton and PI lepton production at NLO and LO.
The contribution from PI process obtained with the photon PDF is scaled with the factor of 100 for better visibility. The fraction of PI contribution in dilepton mass spectrum is generally less than 1\%.
\end{tolerant}

\begin{figure}
\begin{center} 
\includegraphics[width=0.42\textwidth]{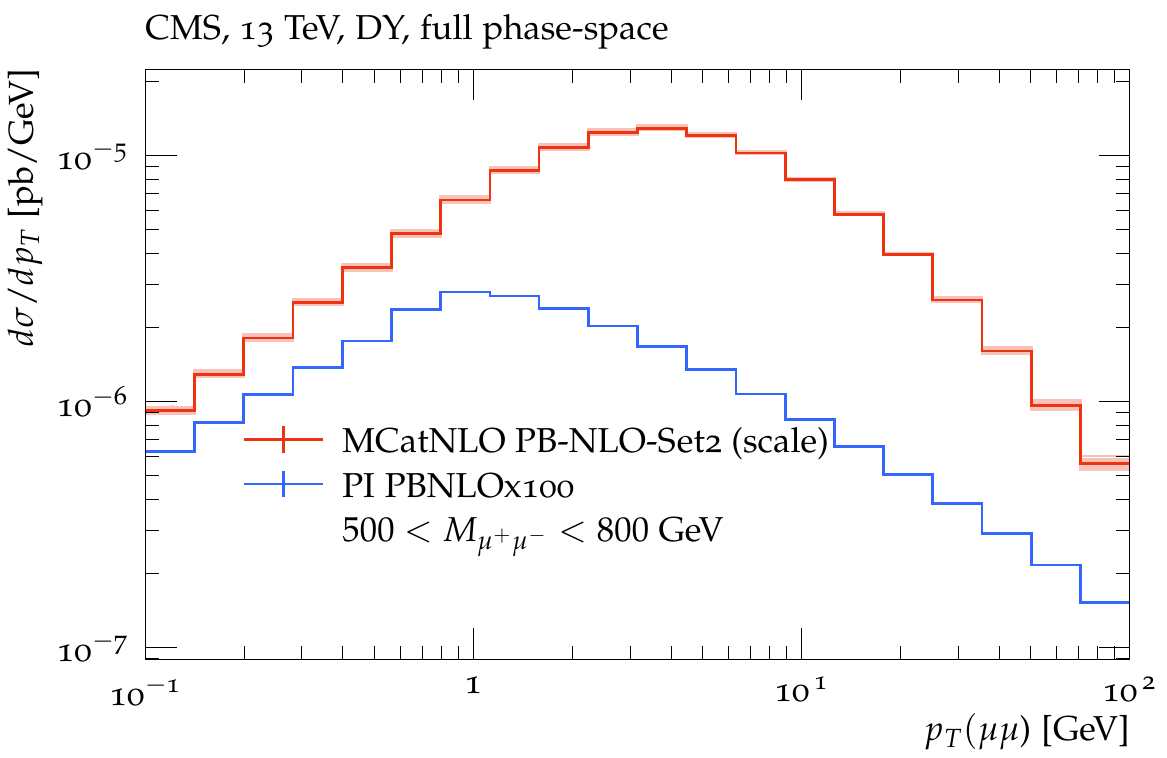}
\includegraphics[width=0.42\textwidth]{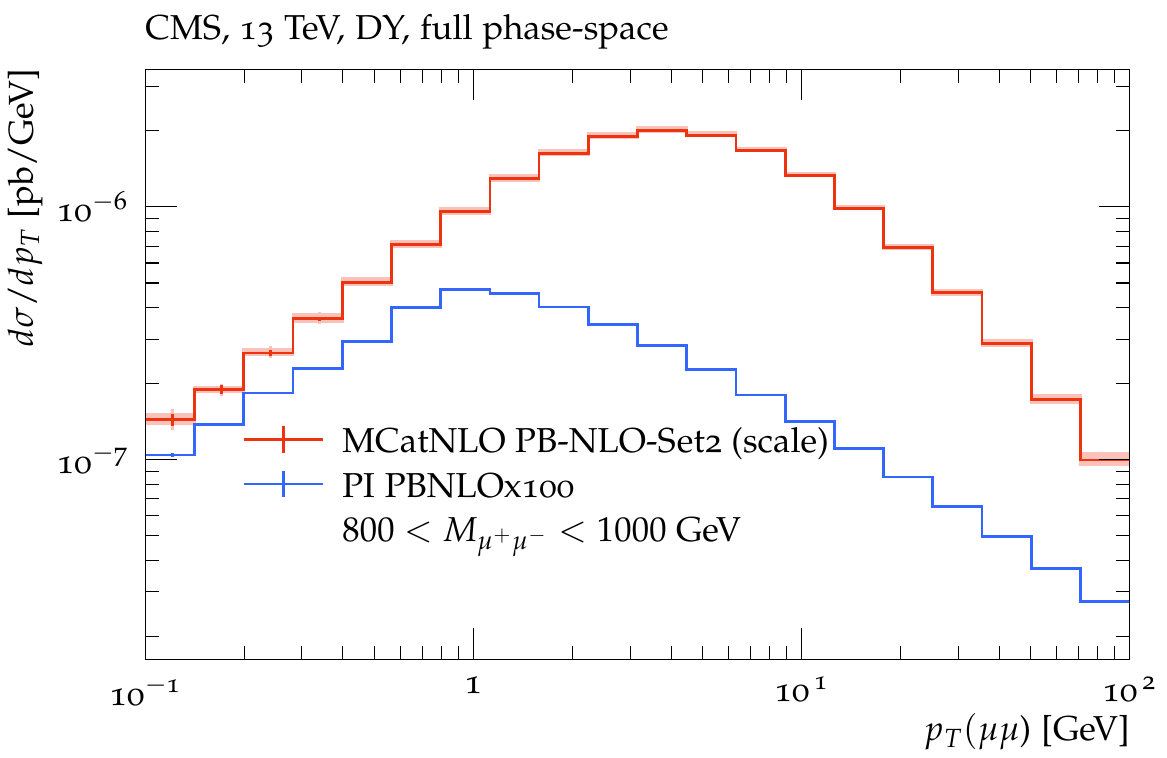}
\includegraphics[width=0.42\textwidth]{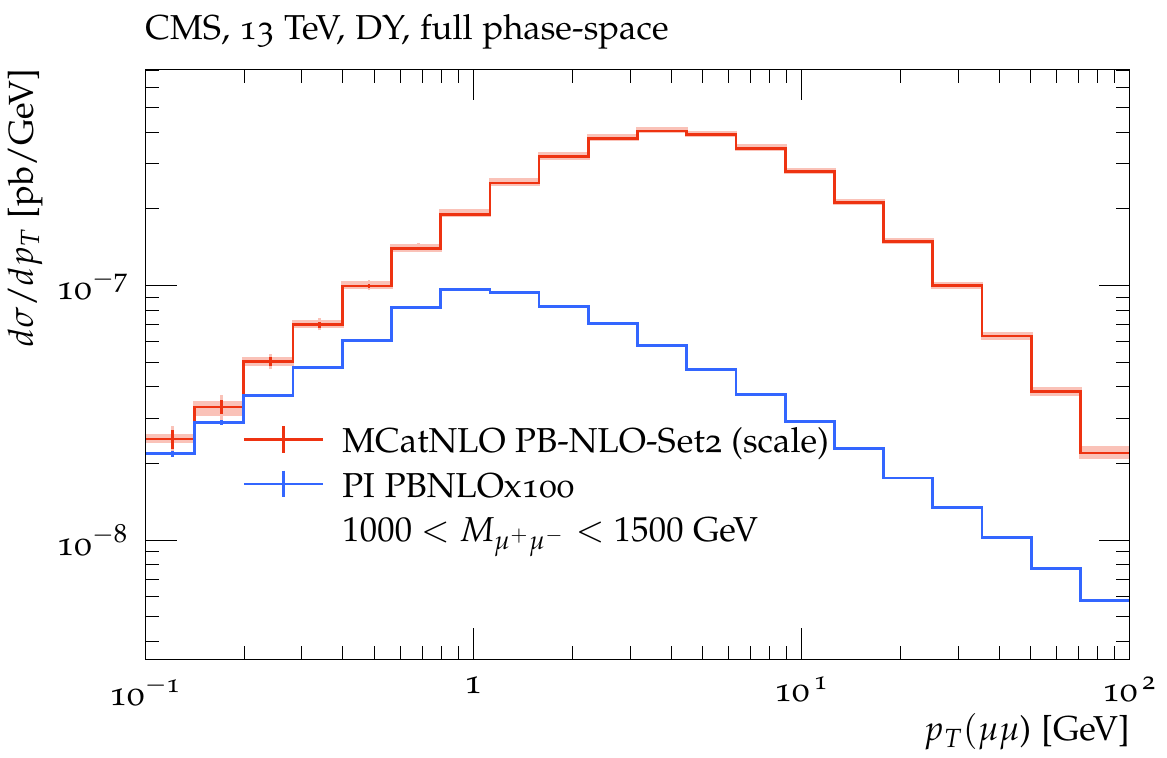}
\includegraphics[width=0.42\textwidth]{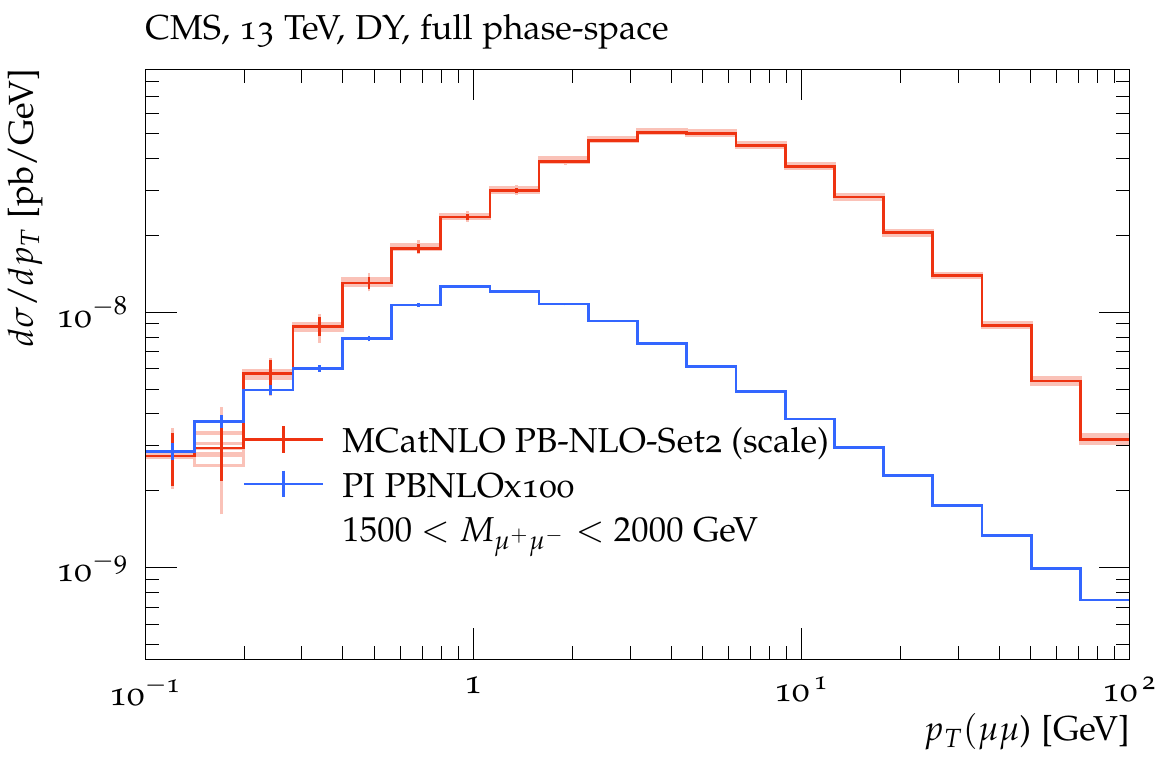}
  \caption{Standard DY and PI transverse momentum spectra based on collinear and TMD PB-QED (Set2) at different high mass regions.}
\label{TMD_pdfs4}
\end{center}
\end{figure}

In Fig. \ref{TMD_pdfs4} we present predictions from NLO PB-TMD-QED and NLO matrix elements for Drell-Yan (DY) transverse momentum spectra for different lepton pair mass regions performed with the \cascade 3 MC generator package ~\cite{Baranov:2021uol}. 
 We also calculate the contribution of PI processes in the transverse momentum spectrum of very high mass dileptons with a collinear and TMD photon density. As shown in Fig. \ref{TMD_pdfs4} for different mass regions, the transverse momentum spectrum of very high DY mass is different between the standard DY and PI lepton spectra. 
 The difference comes from the hard matrix element process, rather than from the TMD distribution.
\section{Conclusion}
We determined collinear and TMD photon densities with the PB method and investigated the mass and transverse momentum spectra of DY lepton-pair production at very high DY masses by matching PB-QED (Set~2) distributions to NLO calculations via MC@NLO. We observed a good description of the dilepton mass measurements in the range 15 to 3000 GeV at $\sqrt{s}=13$ TeV.  We also provided a new perspective on the contribution of PI lepton process in the  transverse momentum spectrum of very high mass lepton pairs.
 
Extracting the photon PB TMD lays the ground work needed to generate collinear PDFs and TMDs for the heavy gauge bosons $Z$, $W$ by implementing the EW sector.
\section*{Acknowledgments.}
\begin{tolerant}{1500}
We thank F. Hautmann for various discussions and comments on the manuscript.
STM thanks the Humboldt Foundation for the Georg Forster research fellowship  and 
gratefully acknowledges support from IPM. 
\end{tolerant}

\providecommand{\href}[2]{#2}\begingroup\raggedright\endgroup

\end{document}